\documentclass[twocolumn,showpacs,preprintnumbers,amsmath,amssymb,letter]{revtex4}

\usepackage{graphicx}
\usepackage{dcolumn}
\usepackage{bm}
\usepackage{subfigure}

\begin{document}

\title{Onset of Patterns in an Ocillated Granular Layer:
  Continuum and Molecular Dynamics Simulations}

\author{J. Bougie, J. Kreft, J. B. Swift, and Harry L. Swinney}
 \affiliation{Center for Nonlinear Dynamics and Department of Physics,
{University of Texas at Austin, Austin, TX 78712}}

\date{\today}

\begin{abstract}
We study the onset of patterns in vertically oscillated
layers of frictionless dissipative particles.  Using both
numerical solutions of
continuum equations to Navier-Stokes order and molecular dynamics (MD)
simulations, we find that standing
waves form stripe patterns above a critical acceleration of the cell.
Changing
the frequency of
oscillation of the cell changes the wavelength of the resulting pattern;
MD and continuum simulations both yield wavelengths in accord with
previous experimental results.  The value of
the critical acceleration for ordered standing waves is
approximately 10\% higher in molecular dynamics simulations than in 
the continuum simulations, and the amplitude of the waves differs significantly
between the models.  The delay in the onset of order in molecular
dynamics simulations and the amplitude
of noise below this onset are consistent with the presence of
fluctuations which are absent in the continuum theory.  The
strength of the noise obtained by fit to Swift-Hohenberg theory
is orders of magnitude larger than the thermal noise
in fluid convection experiments, and is comparable to the noise found in
experiments with oscillated granular layers and in recent fluid
experiments on fluids near the critical point.  Good agreement is found
between the mean field value of onset from the Swift-Hohenberg fit and the onset in
continuum simulations.  Patterns are compared in
cells oscillated at two different frequencies in MD; the layer with larger
wavelength patterns has less noise than the layer with smaller
wavelength patterns.
\end{abstract}

\pacs{45.70.Qj,05.40.Ca,47.54.+r}

\maketitle

\section{Introduction}

\subsection{Background}

A successful hydrodynamic theory of granular media
could allow scientists and engineers to exploit the powerful techniques of
fluid dynamics to describe granular phenomena.  
Recent experiments \cite{bocquet, rericha} and simulations \cite{ramirez}
demonstrate the potential for hydrodynamic theory to describe granular
media; however, the validity of such methods has not yet been
established for a general description of granular flow phenomena
\cite{dufty2002, campbell,rericha2004}.

Several proposed rapid granular flow models use equations of motion
for continuum fields -- number density $n$, velocity ${\bf u}$, and
granular temperature $T$ (${3\over2} T$ is the average kinetic energy due
to random particle motion)  \cite{haff,jenkinsandsavage, lun1984}.  In one
approach, particle interactions are modeled with binary, inelastic
hard-sphere collision operators in kinetic theory to derive continuum
equations to Euler \cite{goldshtein1}, Navier-Stokes
\cite{jenkinsandrichman}, and Burnett \cite{selaandgoldhirsch} order.  In
this paper, we use 3D simulations of continuum equations to
Navier-Stokes order and 3D inelastic hard-sphere molecular dynamics
(MD) simulations to investigate the onset of standing wave patterns in
vertically oscillated granular layers.

\subsection{Standing wave patterns in oscillated granular layers}

Vertically oscillated layers have provided an important testbed for
granular research.  Flat layers of grains on a
plate oscillating sinusoidally in the direction of gravity exhibit
convection \cite{knight96},
clustering \cite{falcon99}, shocks \cite{goldshtein2}, steady-state
flow fields far from the plate \cite{brey01}, and standing wave pattern
formation \cite{melo}.

A layer of grains on a plate oscillating sinusoidally in the direction of
gravity with frequency $f$ and amplitude $A$ will leave the plate at some
time in the cycle if the maximum
acceleration of the plate is greater than that of gravity.  The layer
dilates above the plate, then collides with the plate later in the cycle
and is compressed on the plate by this collision.  Above a critical value of
acceleration, standing wave patterns spontaneously form in the layer.  This
pattern is subharmonic with respect to the plate, repeating every $2/f$
\cite{melo}.

Various subharmonic standing wave patterns, including stripe, square, and
hexagonal patterns, have been found experimentally,
depending on the nondimensional frequency $f^{*}=f\sqrt{H/g}$ and the
nondimensional accelerational amplitude $\Gamma=A\left(2\pi f\right)^2/g$,
where $H$ is the depth of the layer as poured, and $g$ is the acceleration
due to gravity \cite{melo}.  

Studies using hydrodynamic equations have not yet yielded
the standing wave patterns observed in experiments.  Here we investigate the
onset of ordered standing wave patterns
using fully three-dimensional (3D) simulations of continuum equations to
Navier-Stokes order as well as molecular dynamics (MD) simulations.
We use a continuum model for frictionless, inelastic
particles, and
investigate the onset of stripe patterns.

\subsection{Fluctuating hydrodynamics}

Near the onset of convection patterns in Rayleigh-B\'{e}nard convection of
fluids, fluctuations caused by thermal noise create deviations from
dynamics predicted from linear theory.  These fluctuations are described by
the addition of terms to the Navier-Stokes equations; this theory is known
as fluctuating hydrodynamics
\cite{landauandlifshitz1959,zaitsev,swifthohenberg}.  Recent experiments
have shown that fluctuating
hydrodynamics theory accurately describes the dynamics of fluids near the
onset of convection \cite{wu, rehberg, oh}.

Experimental investigations of coherent
fluctuations and pattern formation in oscillated granular layers have indicated
that fluctuations due to the movement of individual grains play a much more
significant role in the collective behavior of granular media than do
thermal fluctuations in ordinary fluids \cite{goldman2004}.  Thus, a
consistent theory of granular hydrodynamics may need to include fluctuations.

\subsection{Model system}
We simulate a layer of grains on an impenetrable plate which oscillates
sinusoidally in the direction of gravity.
The layer depth at rest is approximately $H=5.4\sigma,$ where the grains
are modeled as identical,
frictionless spheres with diameter $\sigma$ and coefficient of restitution
$e=0.7$.   For most of the paper, we study the onset of patterns as a function of 
$\Gamma$, while the frequency of
plate oscillation is held constant at $f^{*}=0.4174$.
This corresponds to
a frequency of $56$ Hz for particles with a diameter of $0.1$
mm. 
For $\Gamma\gtrsim2.5$, stripes are seen experimentally for
a range of parameters, including $f^{*}=0.4174$, $H=5.4$ \cite{melo}.  
In
Sec.~\ref{section-dispersion} and Sec.~\ref{section-wavelength}, frequency
is varied to
investigate the effect of changing frequency on pattern formation.  

Experiments \cite{goldman03} and MD simulations \cite{moon03} indicate that
inter-particle friction plays
an important
role in the standing wave patterns.  MD simulations with friction between
particles have quantitatively reproduced the stripe, square, and hexagonal
subharmonic standing wave patterns seen experimentally for a wide range of
parameters \cite{bizon98}.
However, MD simulations using frictionless particles do not yield stable
square or
hexagonal patterns, but only yield stripe patterns, and exhibit the onset
of patterns at lower $\Gamma$ than that seen for frictional particles
\cite{moon03}.  This result is consistent with experiments which show that
reducing
friction by adding
graphite can de-stabilize square patterns \cite{goldman03}.  In
this study, we neglect the effects of friction in our continuum and MD
simulations, and study only the onset of stripe patterns in frictionless layers.
To
investigate other patterns such as squares or hexagons, simulations would
need to include friction between particles.
 
We use MD
and continuum
simulations to investigate the dynamics of this system near onset, and use
simulations of the Swift-Hohenberg (SH) model equation to compare our
results between the two.  Section II describes the
methods used to simulate and analyze these patterns, Sec. III compares
patterns formed in MD and continuum simulations.
Section IV compares
MD simulations to Swift-Hohenberg theory, and Sec. V presents
our conclusions.
 
\section{Methods}
\subsection{Molecular dynamics simulation}

We use an inelastic hard sphere molecular dynamics simulation, 
which was previously used in conjunction with the continuum
simulation used in this paper to model shock waves in a
granular shaker \cite{bougie2002}.
This same MD code with friction added has been found to describe well the
patterns observed in experiments on oscillating granular layers
\cite{bizon98,moon02}.

The collision model assumes instantaneous binary collisions in which
energy is dissipated, as characterized by the normal coefficient of
restitution $e$.  We neglect surface
friction between particles, as well as between the particles and the
plate.  To prevent inelastic collapse, we use a coefficient of restitution
which depends on the relative colliding velocity of the particles $v_n$:
$e\left( v_n\right) = 1-0.3\left( v_n/
\sqrt{g\sigma}\right)^{3/4}$ for $v_n < \sqrt{g\sigma}$, and
$e=0.7$ otherwise \cite{bizon98}.  

The MD simulations are calculated in a box
of size $L_x \times L_y$ in
the horizontal directions $x$ and $y$, where $L_x$ and $L_y$ are varied to
investigate patterns with different wavelengths.
The simulations use periodic boundary
conditions in the horizontal directions, an impenetrable lower plate
which oscillates sinusoidally between $z=0$ and $z=2A$, and an
upper plate fixed at a height $z=200\sigma$, as in the previous
investigation of shock propagation \cite{bougie2002}.

For each MD simulation, $\left(L_x/\sigma\right)
\times \left(L_y/\sigma\right) \times6$ particles
were used.  In actual packings seen experimentally, $6/\sigma^2$ particles
per unit area
of the bottom plate corresponds to a layer depth $H=5.4\sigma$ as poured,
representing a volume fraction  $\nu\approx0.58$.
\cite{bizon98}. The
total mass of the layer matches that of the continuum
simulations.

\subsection{Continuum simulation}

We use a continuum simulation previously used to model shock waves in a
granular shaker \cite{bougie2002}.  Our simulation numerically integrates
continuum equations of Navier-Stokes order proposed by Jenkins and Richman
\cite{jenkinsandrichman} for a dense gas composed of frictionless
(smooth), inelastic hard spheres.  We integrate these
hydrodynamic equations to find number density, momentum, and granular
temperature, using a second order finite difference scheme on a uniform
grid in 3D with first order adaptive time stepping \cite{bougie2002}.

As in our MD simulations, the granular fluid in the continuum simulations
is contained between two
impenetrable horizontal plates at the top and bottom of the container,
where the lower plate oscillates sinusoidally between height $z=0$ and
$z=2A$.  In our MD simulations, the ceiling is fixed in space
at a height of $z=200\sigma$, but to minimize computation time, the
ceiling in continuum simulations is located at height $80 \sigma$ above the
lower plate and oscillates with the bottom plate.
In our previous study of shock formation,
changing the ceiling height from $200 \sigma$ to $80 \sigma$ resulted in a
fractional root mean square difference of less than $1 \%$ in the shock
location over one cycle \cite{bougie2002}.

As in the MD simulations, we use periodic horizontal boundary conditions
and boxes of size $L_x \times L_y$ in
the horizontal directions $x$ and $y$, where $L_x$ and $L_y$ are varied.
In each case, continuum simulations are compared to MD simulations with the
same horizontal dimensions $L_x$ and $L_y$. 
The numerical methods, boundary
conditions at the top and bottom plate, and grid spacing are the same as
used in the previous study of shocks~\cite{bougie2002}.

The energy loss due to collisions in continuum
simulations is characterized by a single parameter, the normal coefficient
of restitution $e=0.70$.  Throughout this paper, we use units such that
particles in MD simulations have mass unity, and the total mass of the
layer in the continuum simulations matches that used in MD simulations.

\subsection{Characterizing patterns}\rm\label{section-characterizing}

To visualize peaks and valleys formed by standing wave patterns, we calculate
the height of the center of mass of the layer,  $z_{cm}\left( x, y,
t\right)$ as a function of
horizontal
location in the cell at various times in the cycle.  At a given time $t_0$
and horizontal location $\left( x_0, y_0
\right)$, $z_{cm}\left(x_0, y_0, t_0\right)$ is the
center of mass of all particles whose horizontal coordinates lie within a bin
of size $\Delta x_{bin} \times \Delta y_{bin}$ centered at $\left( x_0, y_0
\right)$.  For
continuum simulations, we use the simulation grid size to define the bins:
$\Delta x_{bin}
=\Delta x=2\sigma$ and $\Delta y_{bin}=\Delta y=2\sigma$.  For MD
simulations, we use bins
of size
$2\sigma\times2\sigma$ in Section~\ref{section-patterns} to compare to
continuum simulations with the same bin size.  Peaks in the
pattern correspond to maxima of
$z_{cm}$, and valleys correspond to minima.

To measure the amplitude
of patterns and fluctuations in continuum and MD simulations, we examine
the deviation of the height of the
center of mass of the layer as a function of horizontal location
in the
cell from the center of mass height averaged over the cell at that phase in
the cycle:
\begin{equation}\psi(x,y,t)=z_{cm}(x,y,t)-\left< z_{cm}(x,y,t)
  \right>,\end{equation} where $x$ and $y$ are the
horizontal coordinates, $t$ is the time in the cycle,  $z_{cm}(x,y)$ is the
height of the center of mass of the layer at horizontal location $(x,y) $,
and the
brackets represent an average over all horizontal locations in the cell at
a given time $t$.

Throughout this paper, we characterize the patterns at the beginning of a
sinusoidal oscillation cycle, such that the plate is at its equilibrium
position and moving upwards.
Using this definition, $\left< \psi^2(t) \right>$ represents the mean square
deviation of the height of the layer from the mean height of the layer at
that phase of the plate.
We note that $\left< \psi^2 \right>$ is large for layers with high
amplitude
patterns or fluctuations, and goes to zero as the layer becomes perfectly flat.

In addition to $\left< \psi^2 \right>$, we
distinguish between ordered patterns (stripes) and disordered fluctuations
by characterizing the long range order of the pattern.
To characterize the long range order of the
patterns, we first calculate the power spectrum of the pattern: 
$S\left(k_x,k_y,t\right)=\left|\tilde{\psi}\left(k_x,k_y,t\right)\right|^2$,
where
$\tilde{\psi}\left(k_x,k_y,t\right)=\int_{0}^{L_x}\int_{0}^{L_y}\psi\left(x,y,t\right)e^{-ik_x
  x}e^{-ik_y y}dx dy$.  We then transform to polar coordinates in $k$ space:
$k_r=\sqrt{k_x^2+k_y^2}$, $k_{\theta}=tan^{-1}\left(k_y/k_x\right)$ to find 
$S(k_r,k_{\theta})$ in the range  $0\leq k_{\theta} < \pi$. 
We integrate radially to find the
angular orientation of the power spectrum: 
$S(k_{\theta})=\int_0^{K}
S\left(k_r,k_{\theta}\right) k_r dk_r,$ where $K=\frac{2\pi\Delta x_{bin}}{L_x}$.
We bin $k_{\theta}$ into 21 bins
between $k_{\theta}=0$ and $k_{\theta}=\pi$, and characterize the long range
order of the patterns by the fraction of the total integrated power that
lies in the bin
with the maximum power:
\begin{equation}P_{max}={S_{max}\left(\theta\right)\over\int_{0}^{\pi}S\left(\theta_{i}\right)dk_{\theta}},\label{eq:order}\end{equation}
where $S_{max}\left(\theta\right)$
is the integrated power within an angle $\pi/21$ of the maximum value of
$S(\theta)$.  Thus $P_{max}$ is the fraction of the total power that lies within
approximately $\pi/21$ of the angular location of the maximum power.  For a 
perfectly disordered state, with equal power in all
directions, $P_{max}$ would approach ${1\over 21} \approx 0.05$, while 
$P_{max}=1$ for a state with
all power in a single bin.  
Thus $P_{max}$
provides a measure of order when stripes form.

\section{Pattern onset and dispersion}\label{section-patterns}
\subsection{Stripe patterns}\label{section-bigover}

Experimental investigations of shaken granular
layers have shown that above a critical acceleration of the plate
$\Gamma_c$, standing wave patterns form spontaneously.  These patterns
oscillate subharmonically, repeating every $2/f$, so that the location of
a peak of the pattern becomes a valley after one cycle of the plate, and
vice versa \cite{melo}.  

\begin{figure}

\subfigure{\label{overmd}\scalebox{.3}{\includegraphics{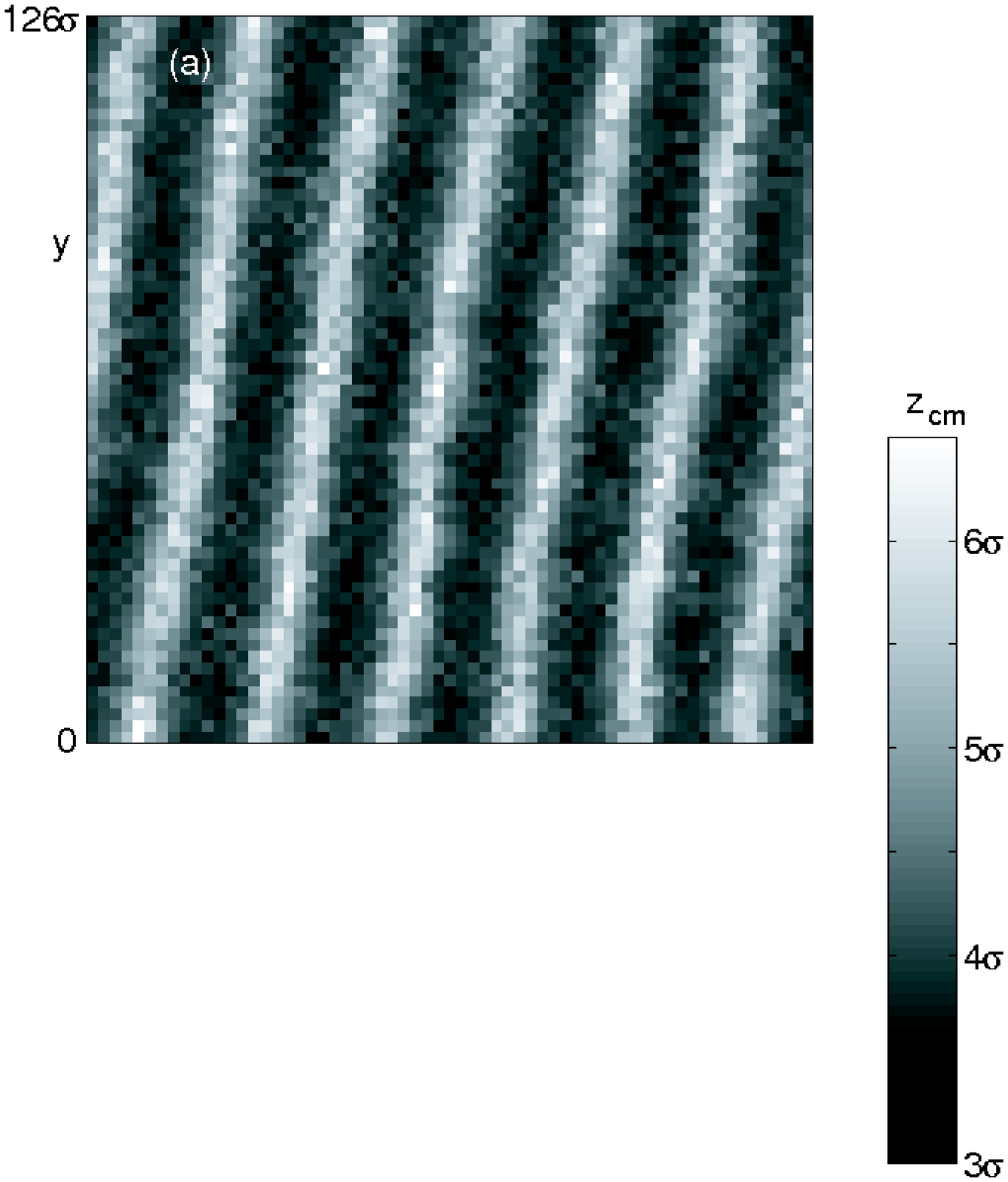}}}\\
\vspace{-2.cm}
\hspace{-1.1cm}
\subfigure{\label{overcont}\scalebox{.3}{\includegraphics{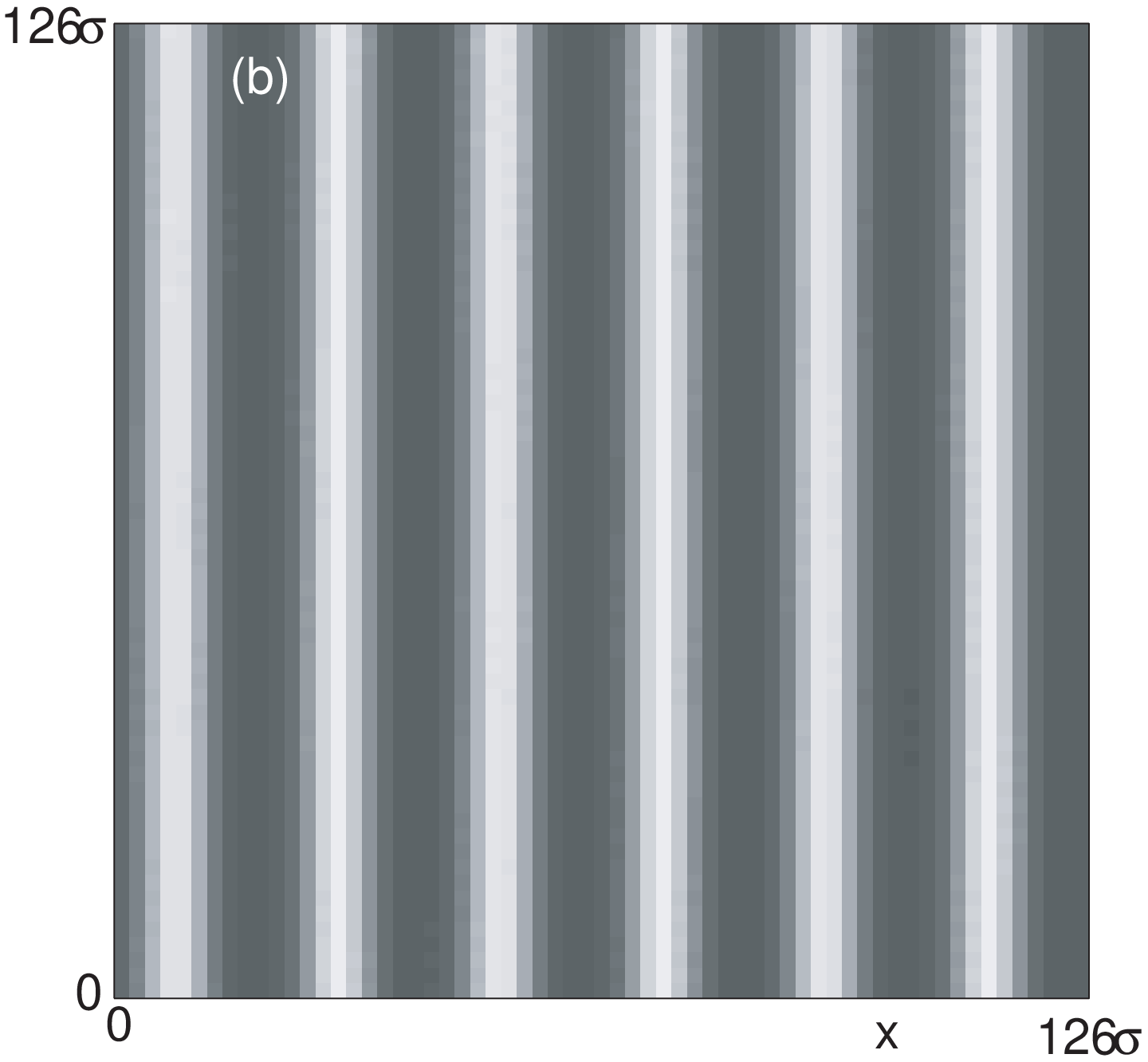}}}\\
\caption{\label{overpic} An overhead view of a layer of grains, showing the
  center of mass height $z_{cm}$ as a function of horizontal position
  $\left( x,y\right)$ in a cell with horizontal dimensions $L_x \times L_y = 126\sigma
  \times 126\sigma$, from (a) MD simulations and (b) continuum
  simulations.  Peaks of the layer corresponding to large center of mass
  height $z_{cm}$ are shown in white; valleys corresponding to low
  $z_{cm}$ are shown in black.
}
\end{figure}

Continuum and MD simulations produce standing wave patterns for
$\Gamma=2.2$ and
$f^{*}=0.4174$ (Fig.~\ref{overpic}).  Alternating peaks and valleys form a
stripe
pattern which oscillates at $f/2$ with respect to the plate oscillation; a
location in the cell which represents a peak during one cycle will become a
valley the next cycle, and then return to a peak on the following cycle.
For a box of size
$126 \sigma \times 126 \sigma$ in the horizontal direction, six wavelengths
fit in the box in both MD and continuum simulations, yielding a wavelength
of $21\sigma \pm 4\sigma$ in both continuum and MD simulations
(Fig.~\ref{overpic})
.

\subsection{Dispersion Relations in Continuum, MD, and Experiment}\label{section-dispersion}

Experiments have shown that the wavelength $\lambda$ of standing
wave patterns in
shaken granular layers depends on the frequency of the plate oscillation
\cite{melo1993, clement1996, umbanhowar2000}.  For a range of layer depths
and oscillation frequencies, experimental data for
frictional particles near the onset of patterns were found to be fit by the
function $\lambda^{*}=1.0+1.1f^{*-1.32\pm0.03}$, where
$\lambda^{*}=\lambda/H$ \cite{umbanhowar2000}.

We investigate the frequency dependence of standing waves in continuum
simulations and in MD simulations of frictionless particles.  
Dimensionless accelerational amplitude $\Gamma=2.2$ was held constant
while dimensionless frequency $f^*$ was varied.  Simulations
were conducted in a box of horizontal extent $L_x=168\sigma$ and
$L_y=10\sigma$.  This orientation causes stripe patterns to form
parallel to the $y-$ axis.  The
dominant wavelength in each case was calculated from $S\left(
k_x,k_y,t\right)$ by finding the wavenumber $k_x$
in the $x-$ direction which exhibited the maximum power during
one cycle of the oscillatory state of the pattern.  Due to the
periodic boundary conditions and finite box size, wavelengths must
fit in the box an integer number of times.  This finite size
effect of quantized wavelength yields
inherent uncertainty in the wavelength that would be selected in an infinite
box.  

\begin{figure}
\begin{center}

\scalebox{.5}{\includegraphics{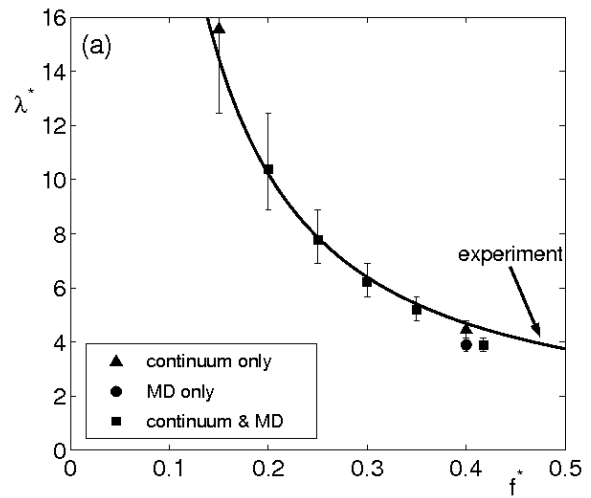}}\\
\end{center}

\vspace{0cm}

\caption{\label{dispersion} 
Dispersion relation for
stripes which form perpendicular to the long dimension of 
cells with horizontal dimensions $168\sigma\times10\sigma$.  Data for
continuum simulations are shown as triangles and MD simulations as circles;
points where continuum and MD simulations yield the same wavelength are
shown as squares.  
In both continuum
and MD simulations, the dominant
wavelength of the final oscillatory state $\lambda$ fits very well to the
dispersion relation found in experiments 
$\lambda^{*}=1.0+1.1f^{-1.32\pm0.03}$ (solid line) \cite{umbanhowar2000}.
Error bars in both
simulations are calculated exclusively from discretization due to
periodic boundary conditions in a finite size box.
}
\end{figure}

Wavelengths found in continuum and MD simulations are compared to the
dispersion relation fit to experimental data in Fig.~\ref{dispersion}.
Investigation is limited to 
$f^{*}>0.15$ by the box size, as only two wavelengths fit in
the box in continuum simulations at this frequency.  Neither simulation
produced patterns for this box size for $f^{*}\gtrsim0.45$.
Both simulations agree quite well with the experimental fit throughout the
range $0.15\lesssim f^{*} \lesssim 0.45$.

Comparison to the
experimental fit shows that both MD and continnum simulations produce
wavelengths consistent with experimental results for frictional particles.
These data
indicate that friction seems to be
unimportant in wavelength selection through this parameter range.  

\subsection{Layers Above and Below the Onset of Patterns}

\begin{figure}

\hspace{-.34in}
\subfigure{\label{GAM1.9md}\scalebox{.27}{\includegraphics{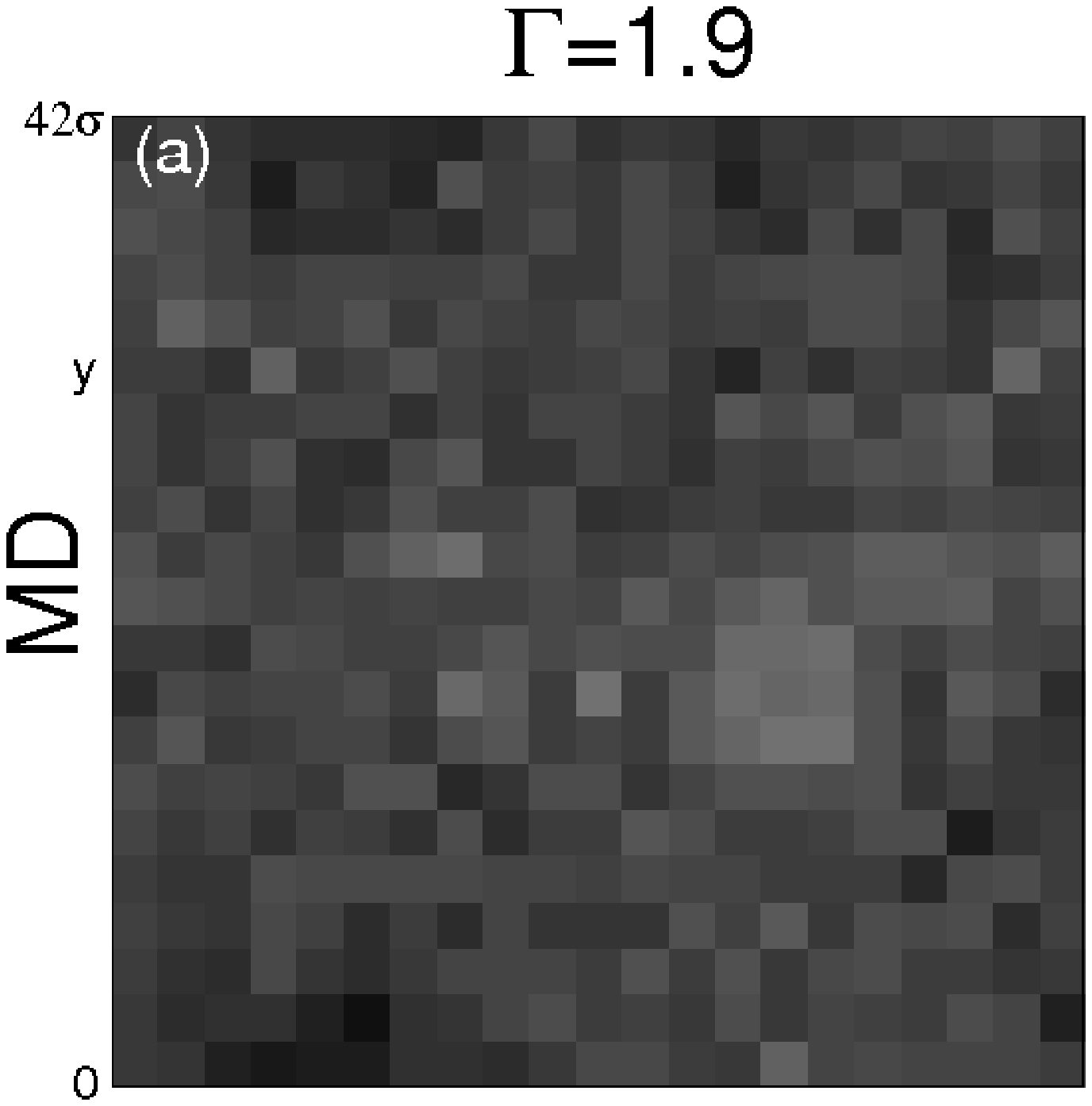}}}
\subfigure{\label{GAM2.2md}\scalebox{.27}{\includegraphics{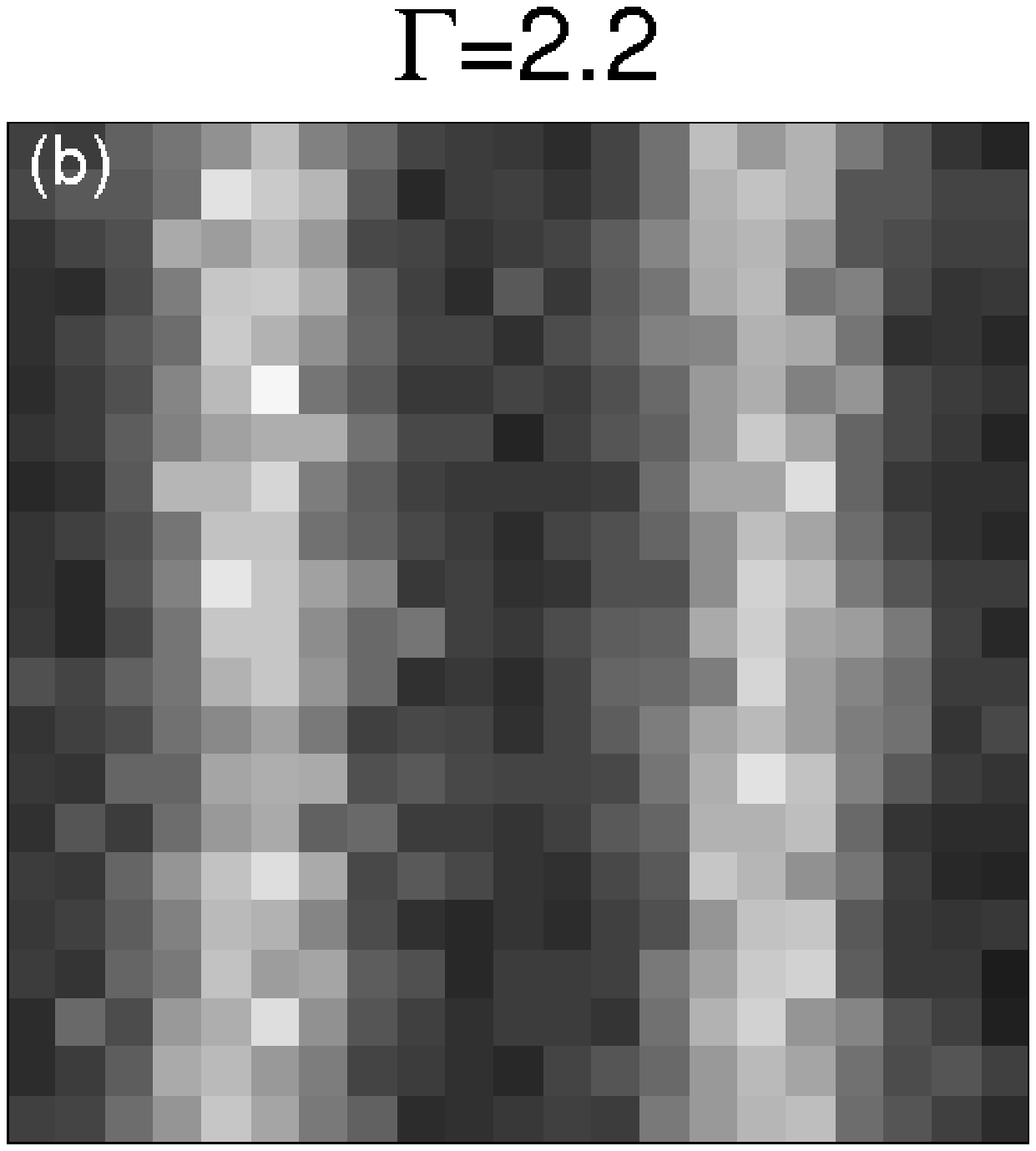}}}\\

\vspace{-2.5cm}
\hspace{7.9cm}
\label{GAMcbar}\scalebox{.27}{\includegraphics{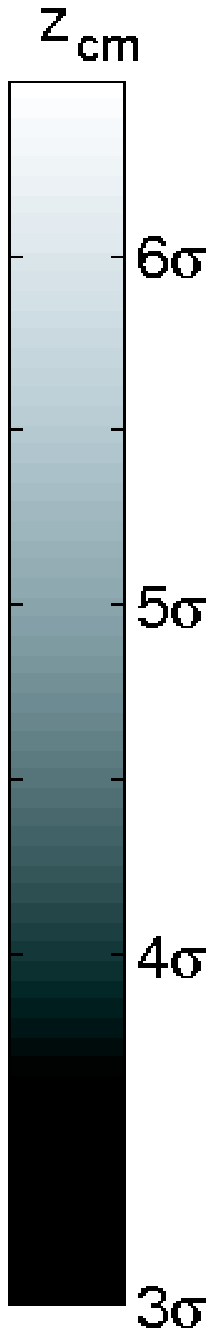}}\\

\vspace{-1.8cm}

\hspace{-.75cm}
\subfigure{\label{GAM1.9cont}\scalebox{.27}{\includegraphics{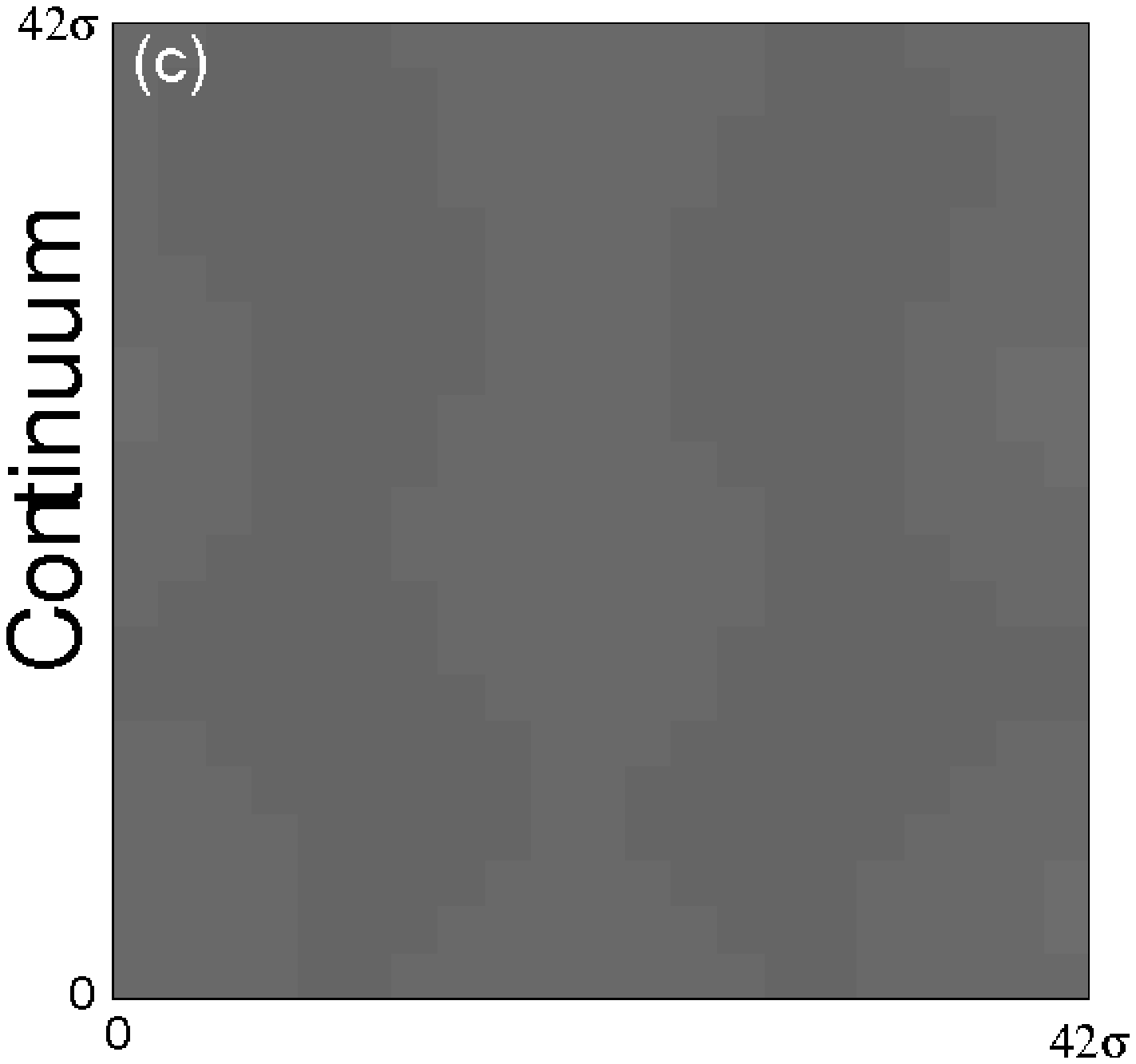}}}
\hspace{.115cm}
\subfigure{\label{GAM2.2cont}\scalebox{.27}{\includegraphics{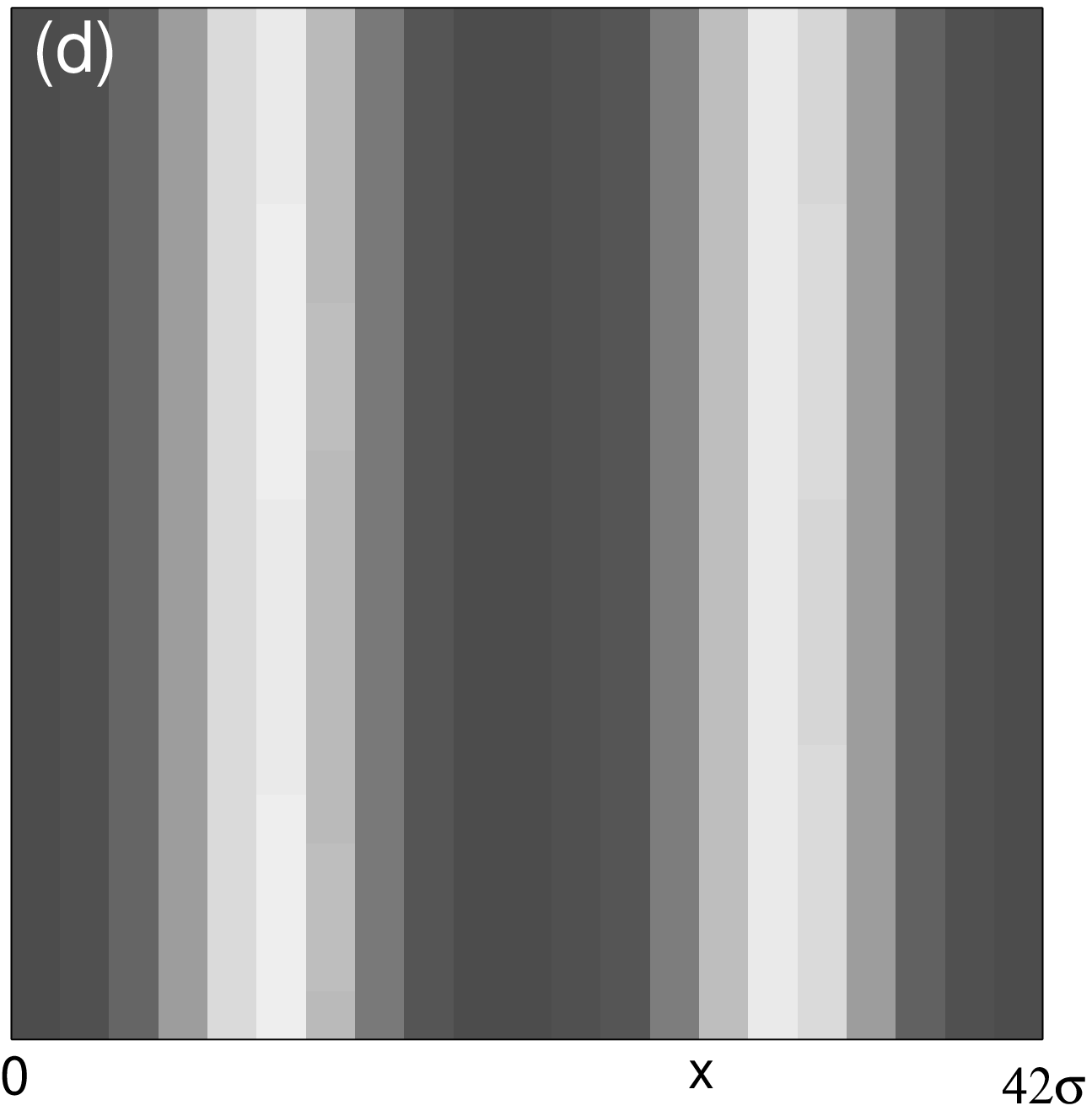}}}

\caption{\label{GAMcomparison} 
An overhead view of the layer of grains, showing the center of mass height
$z_{cm}(x,y)$ of
the layer as a function of location in the box, for (a) MD simulations with a
plate acceleration with respect to gravity $\Gamma=1.9$, (b) MD
simulations with $\Gamma=2.2$, (c) continuum simulations with $\Gamma=1.9$,
and (d) continuum simulations with $\Gamma=2.2$.  Peaks corresponding to
large $z_{cm}$ are shown in white, while valleys corresponding to
small $z_{cm}$ are shown in black.  
The grayscale for all four images is given on the right.
}
\end{figure}

Continuum and MD simulations exhibit pattern formation above a
critical acceleration of the plate; however, standing wave patterns are not
observed below a critical value of $\Gamma$ (Fig.~\ref{GAMcomparison}).  
For $\Gamma=2.2$, both MD
(Fig.~\ref{GAM2.2md}) and continuum (Fig.~\ref{GAM2.2cont}) simulations show well
defined peaks
and valleys which form stripe patterns with two wavelengths fitting in the
box of size $L_x=L_y=42\sigma$.  The only difference between this system and that investigated in
Sec.~\ref{section-bigover} is the horizontal size of the cell; these
patterns look very similar to a section of the patterns formed in the
larger cell
(Fig.~\ref{overpic}).  Reducing the accelerational amplitude to $\Gamma=1.9$
while keeping all other parameters constant yields no ordered waves in
either MD
(Fig.~\ref{GAM1.9md}) or continuum
(Fig.~\ref{GAM1.9cont}).  Thus both continuum and MD simulations appear to
have a critical value of $\Gamma$ somewhere in the range
$1.9\leq\Gamma_c\leq2.2$, such that no patterns are formed for
$\Gamma<\Gamma_c$, and patterns are formed for $\Gamma>\Gamma_c$.  This
critical value is lower than that found in experiments with frictional
particles, where a similar onset of patterns is found at a critical value
of $\Gamma\approx2.5$ \cite{melo}.

Despite the similarities, differences between MD and continuum simulations
are observable.  For  $\Gamma=1.9$, the continuum
simulation yields a very smooth, flat layer (Fig.~\ref{GAM1.9cont}), while MD
exhibits visible fluctuations (Fig.~\ref{GAM1.9md}).  For  $\Gamma=2.2$, the
continuum simulations produce stripes (Fig.~\ref{GAM2.2cont})  which are much
smoother than those found in MD simulation (Fig.~\ref{GAM2.2md}).  

To explore the differences between the two simulations, we investigate
the onset of patterns in
more detail in continuum simulations and MD simulations separately.

\subsection{Onset of patterns in continuum simulations}\rm
We investigate the
onset of patterns in continuum simulations by determining
$\left<\psi^2\right>$ of
standing waves for different values of $\Gamma$.  
Each simulation begins with a flat layer above
the plate with small amplitude random fluctuations.  The simulation is run
until it reaches a periodic state, at which point $\left<\psi^2\right>$ is calculated
as an average over ten cycles of the same phase of the plate.

\begin{figure}
\begin{center}
\scalebox{.5}{\includegraphics{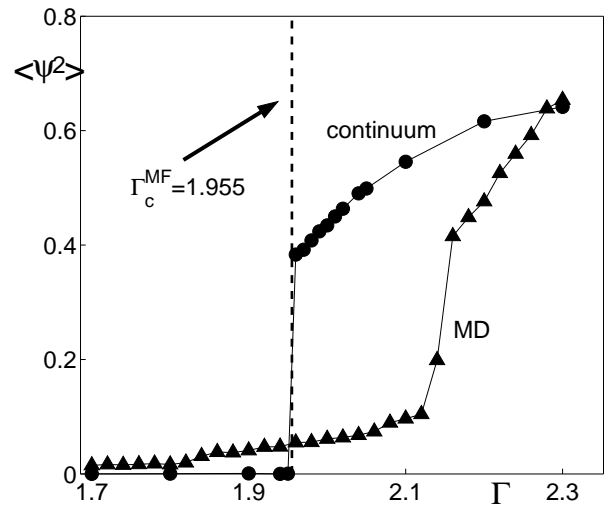}}
\end{center}
\caption{\label{contgrowth}
The mean square deviation $\left< \psi^2 \right>$ of the local center of
mass height from the average center of mass height of the entire layer as a
function of accelerational amplitude $\Gamma$ for MD (triangles) and
continuum (circles) simulations.  The vertical dotted line represents the
onset of stripe patterns in the continuum simulations.
}
\end{figure}

For $\Gamma \lesssim 1.95$, the initial fluctuations decay rapidly until the
layer is quite flat, as represented by
negligible values of $\left<\psi^{2}\right>$ (Fig.~\ref{contgrowth}).  As
$\Gamma$ increases, there is a sudden onset to large amplitude waves, as
seen by the sudden jump in  $\left<\psi^{2}\right>$ in Fig.~\ref{contgrowth}.
This onset occurs at the
critical value $\Gamma_c=1.955\pm0.005$.  For $\Gamma < \Gamma_c$, initial
fluctuations decay until the layer is very flat, while for all layers above
onset ($\Gamma > \Gamma_c$), these waves produce ordered patterns of
stripes similar to those in Fig.~\ref{GAM2.2cont}.

\subsection{Onset of patterns in molecular dynamics simulations}\rm
We examine the onset of
patterns in MD simulations using the same methods as for the continuum
equations.  Figure~\ref{contgrowth} shows the mean square height deviation $\left<
\psi^2 \right>$
as a function of $\Gamma$ for MD simulations as well as for continuum
simulations.  For each value of $\Gamma$, the simulation was run for 400
cycles
of the plate until the layer reached a periodic state, then $\left<\psi^2\right>$ and
$P_{max}$ were calculated from an average of the next 100
cycles.

As in continuum simulations, $\left<\psi^2\right>$ grows
with
increasing $\Gamma$. Unlike the continuum results, $\left<\psi^2\right>$ is
non-negligible in MD simulations even for $\Gamma < 1.95$.  
There is still a sharp increase in the slope of the
curve, but it is delayed until $\Gamma >2.1$.

\section{The role of fluctuations}\rm\label{section-noise}

The MD simulations display
an onset of ordered stripes that is delayed with respect to those found in
continuum, and exhibit non-negligible
$\left<\psi^2\right>$ even
below the onset of ordered stripes.  
Since the
hydrodynamic model used in the continuum simulations does not include a
stochastic noise term characteristic of fluctuating hydrodynamics, the
differences between the continuum and MD simulations may be consistent with
the presence of noise in the MD simulations due to the small number of
particles per wavelength.  To test the hypothesis that these differences
are consistent with the presence of fluctuations in molecular dynamics
simulations, we compare MD
simulations to results from the Swift-Hohenberg model.

\subsection{Swift Hohenberg simulation}

The Swift-Hohenberg (SH) model was
developed to describe thermal noise-driven
phenomena near the onset of long range order in Rayleigh-B\'{e}nard
convection \cite{swifthohenberg}.  
Recent experimental evidence suggests
similar phenomena in shaken granular experiments can
be interpreted using the methods of fluctuating hydrodynamics
\cite{goldman2004}.

The SH model describes the time evolution of a scalar field
$\psi_{SH}(\bf{x}\rm,t)$:
\begin{equation} {\partial\psi_{SH}\over\partial t}=
\left(\epsilon-\left(1+\nabla^2\right)^2\right)\psi_{SH}- \psi_{SH}^3 +
\eta\left(\bf{x}\rm,t\right), \label{eq:SH}\end{equation}
where $\epsilon$ is the bifurcation parameter, and $\eta$ is a stochastic
noise term of strength $F$, such that
$\left<\eta\left(\bf{x}\rm,t\right)
\eta\left(\bf{x'}\rm,t'\right)\right> = 2
F\delta\left(\bf{x}\rm-\bf{x'}\rm\right)\delta\left(t-t'\right)$.
In the absence of stochastic noise ($F=0$), called the mean field (MF)
approximation, there is a sharp onset of stripe patterns
with long range order at $\epsilon=\epsilon^{MF}_c=0$
\cite{swifthohenberg,scherer}.  For $F\neq0$, the
effect of noise is to delay the onset of long
range (LR) order to a new critical value: $\epsilon_c^{LR}>0$.  The delay
in onset is characterized by
$\Delta\epsilon_c=\epsilon_c^{LR}-\epsilon_c^{MF}$.  In
addition, the presence of noise creates fluctuations below the
onset of long range order ($\epsilon<\epsilon_c^{LR}$).  

The Swift-Hohenberg simulation displays a forward bifurcation to stripes at
onset, while MD simulations show slight ($<1\%$) hysteresis
\cite{goldman2004}.  A more complicated SH
model \cite{sakaguchi} yields square patterns
with hysteresis; however, in this work we compare stripe formation in MD
simulations a simpler model of the effects of noise near a bifurcation
(Eq.~\ref{eq:SH}).

We numerically solve the SH equation using the scheme
described in \cite{cross}, with the number of gridpoints $N=42\times 42$,
and
periodic boundary
conditions.  We use integration timesteps of 0.5, and   
the size of each gridspace in the horizontal directions
$\Delta x=\Delta y=0.29$ so that two
wavelengths of the resulting pattern fit in the box, to match MD and continuum
simulations.
The simulation was allowed to run for 8,000
timesteps to reach a final pattern; then $\left<\psi_{SH}^2\right>$ and
$P_{max}$ were calculated from an average of the next 2,000
timesteps, in the same way as $\left<\psi^2\right>$ and $P_{max}$ were
calculated for MD and continuum simulations in
Section~\ref{section-characterizing}.  

\subsection{Comparing Swift-Hohenberg and molecular dynamics simulations}\rm\label{section-fit}

\begin{figure}
\begin{center}

\subfigure{\label{SHamp}\scalebox{.5}{\includegraphics{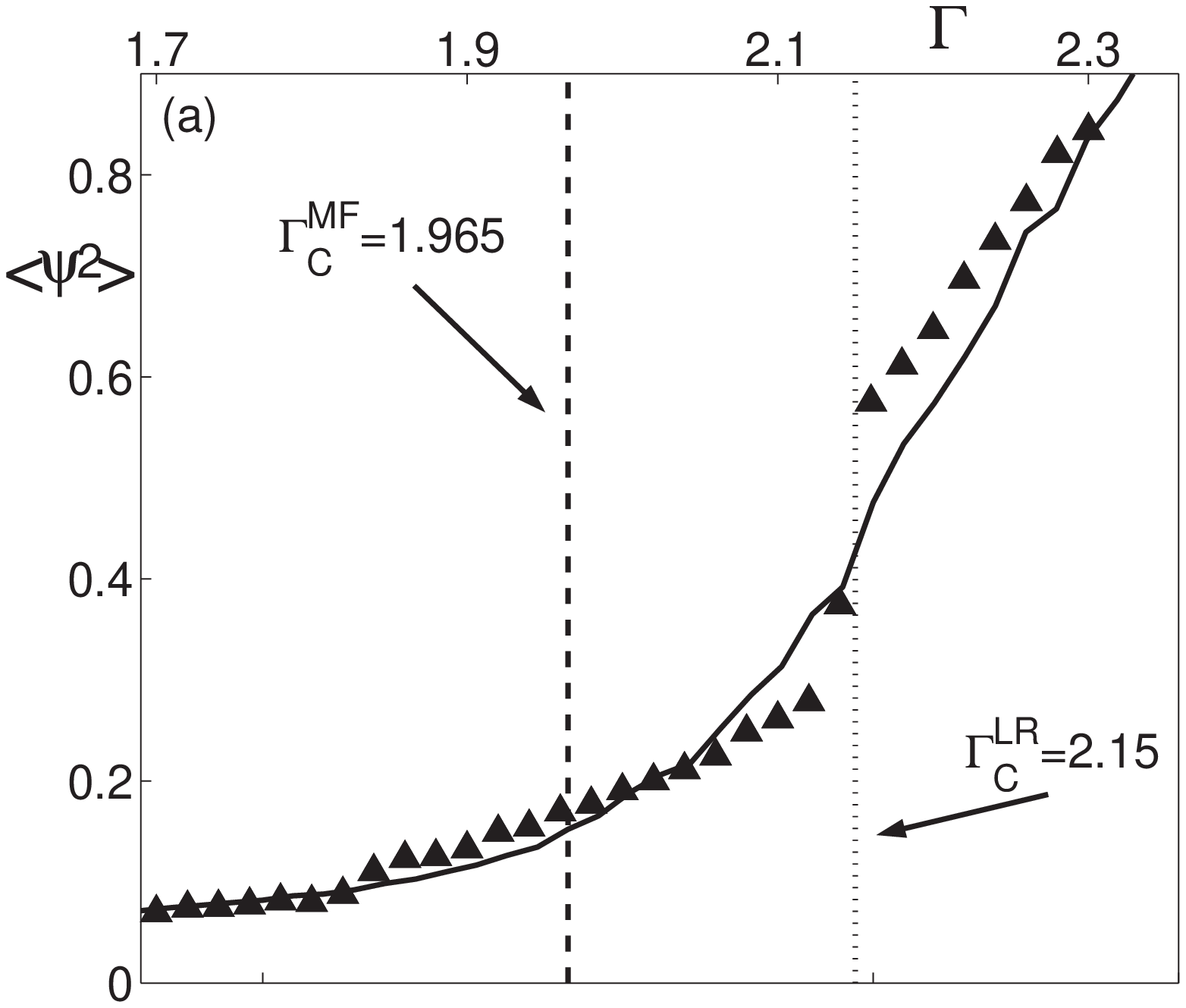}}}\\

\vspace{-.6cm}
\hspace{.5cm}
\subfigure{\label{SHangle}\scalebox{.5}{\includegraphics{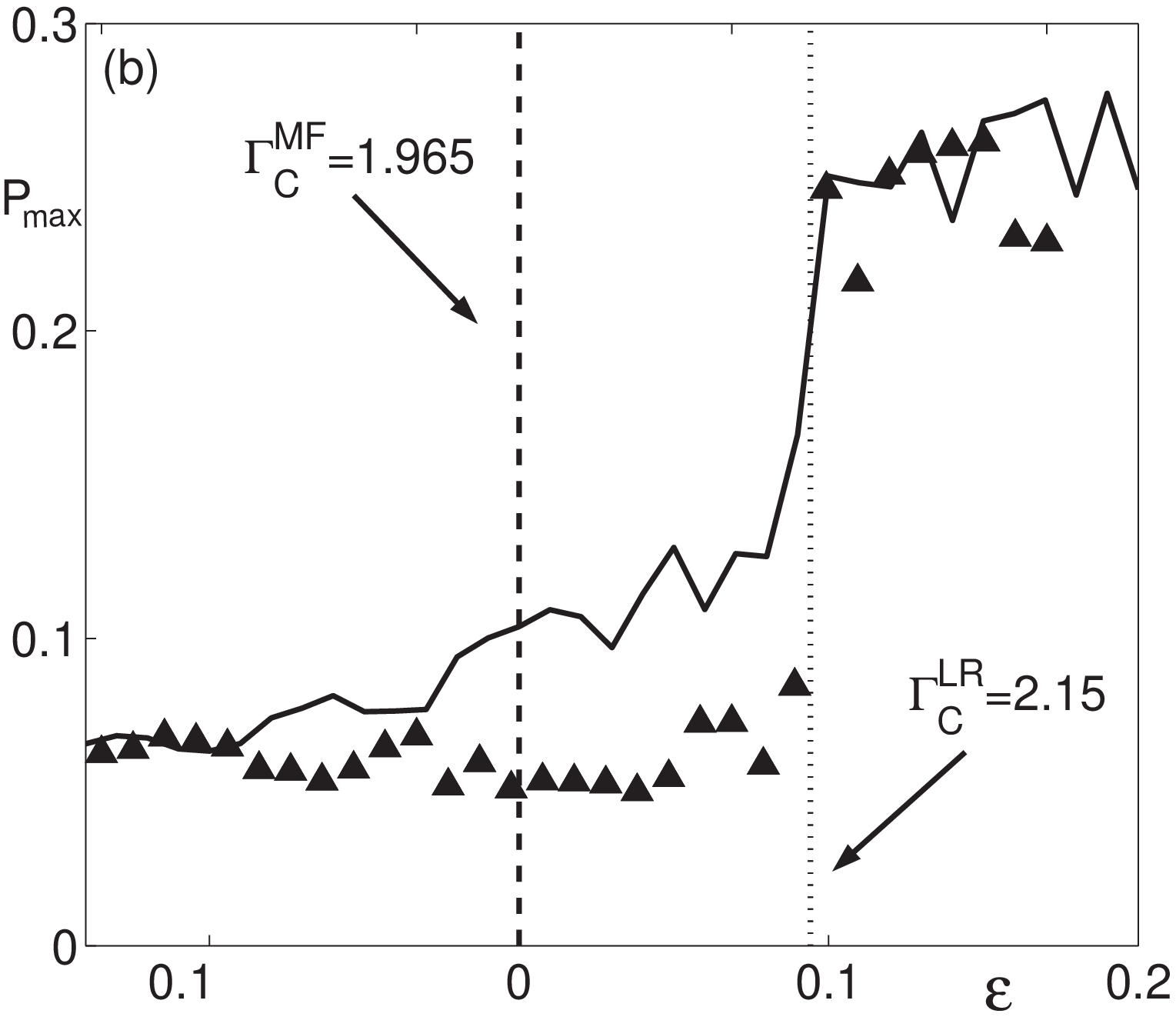}}}
\end{center}
\caption{\label{SHcomp}
Comparison of MD simulations (triangles) to the Swift-Hohenberg model
(solid lines) for (a) $\left< \psi^2 \right>$, and (b) global ordering
$P_{max}$ (Eq.~\ref{eq:order}), as a function of
control parameter $\epsilon$ (bottom axis) for SH, and $\Gamma$ (top axis)
for MD.  The parameters for SH simulations are noise strength
$F=(1.2\pm0.2)\times 10^{-2}$ and a delayed onset of long range
order $\epsilon^{LR}_{c}=0.094$.
The global ordering jumps sharply at
$\epsilon^{LR}_{c}=0.094$, corresponding to $\Gamma_{c}^{LR}=2.15$ in
MD (the vertical dotted line in the figure),
representing a transition to stripe patterns, while
$\left< \psi^2 \right>$ increases
smoothly through that transition.
This fit predicts a mean field onset value of
$\Gamma_{c}^{MF}=1.965\pm0.007$, corresponding to $\epsilon_{c}^{MF}=0$ (the
vertical dashed line in the figure).  
}

\end{figure}

To find the strength of the noise and the mean field onset, 
we fit the SH
model to the data from MD simulations (Fig.~\ref{SHcomp}) by varying three
parameters: $F$, $\Delta\epsilon_c$, and an overall scale
factor, as in \cite{oh, goldman2004}.

Of the three parameters, only the noise strength $F$ changes
the overall shape of the curve.  For a given $F$, the SH simulation is run
for a range of $-0.2\leq\epsilon\leq0.2$; $\psi_{SH}$ and $P_{max}$ are
calculated from the steady state solution for each value of $\epsilon$ and
compared to MD simulations.
For consistency, $\left<\psi^2\right>$ and $P_{max}$ are calculated
for MD simulations from bins of size $\Delta x_{bin}=\Delta
y_{bin}=\sigma$ throughout this section, so that the number
of bins in both MD and SH simulations is $42\times42$.
Increasing the bin size to $\Delta x_{bin}=\Delta
y_{bin}=2\sigma$ does not change any of the fit parameters to within our uncertainty. 

Note $\left<\psi_{SH}^2\right>$ in SH simulations is
found as a function of
control parameter $-0.2\leq\epsilon_{SH}\leq0.2$, while in MD simulations,
$\left<\psi_{MD}^2\right>$ is found as a function of control parameter
$1.7\leq\Gamma\leq2.3$. To compare the onset of the SH model to the onset
in MD simulations, we define
$\epsilon_{MD}=\left(\Gamma-\Gamma_C^{MF}\right)/\Gamma_C^{MF}$, where
$\Gamma_C^{MF}$ is the mean field onset of patterns, comparable to
$\epsilon_{SH}=0$.  However, we do not know {\it a priori} the value of
$\Gamma_C^{MF}$. 

We find that $\left< \psi^2 \right>$ changes relatively
smoothly in MD and SH simulations, making
it difficult to pinpoint an onset of patterns from $\left< \psi^2 \right>$ alone.
However, 
there is a distinct onset of long range order in the system (Fig.~\ref{SHcomp}).
For low $\Gamma$ in MD, the
fluctuations are disordered ({\it cf} \rm  Fig.~\ref{GAM1.9md}), while for higher
$\Gamma$, standing wave stripe patterns are observed ({\it cf} \rm 
Fig.~\ref{GAM2.2md}).  A clear transition from disordered fluctuations to an
ordered stripe pattern is demonstrated by the sharp increase in $P_{max}$ as
$\Gamma$ crosses the critical value for long range order, determined from
Fig.~\ref{SHangle} as $\Gamma_c^{LR}=2.15\pm0.01$.
A similar transition to ordered stripes is seen in SH simulations
(Fig.~\ref{SHangle}).

The onset of long range order is used to establish a correspondence between
$\Gamma$ and $\epsilon$.  For MD simulations, we measure the onset of long
range order as the point of sharpest increase
in $P_{max}$ (Fig.~\ref{SHangle}).  In SH simulations,  $\Delta\epsilon_c$ represents
the onset of long range order.  We match the single point of steepest
increase of $P_{max}$ between the two curves.  The measured value
$\Delta\epsilon_c$ in SH then predicts the mean field onset $\Gamma_C^{MF}$
corresponding to $\epsilon=0$.

Once the relationship between $\Gamma$ and $\epsilon$ is determined, the
overall scale factor for a given $F$ is found by a least squares fit between
$\left<\psi_{SH}^2\right>$ and $\left<\psi_{MD}^2\right>$ for the range
$1.7\leq\Gamma\leq\Gamma_c^{LR}$
(see Fig.~\ref{SHangle}).  This minimization procedure gives the best
possible fit for a given value of $F$.

\begin{figure}
\begin{center}

\scalebox{.5}{\includegraphics{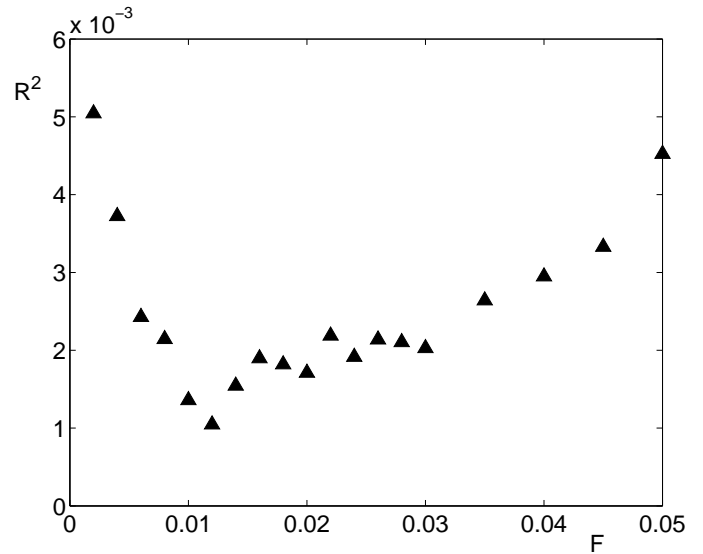}}

\end{center}
\caption{\label{bestfit} 
The squared residual $R^2$ between $\left<\psi_{MD}^2\right>$ and
$\left<\psi_{SH}^2\right>$ as a
function of the noise strength $F$ used in SH simulations.  The best least
squares fit is given by $F=(1.2\pm0.2)\times 10^{-2}$.
}

\end{figure}

This entire procedure is repeated for varying $F$, minimizing the squared
residual 
$R^2=\sum{\left(\left<\psi^2_{MD}\right>-\left<\psi^2_{SH}\right>\right)^2/N},$
 where $N$ is the number of bins (Fig.~\ref{bestfit}).  The best fit yields 
an onset of long range order at $\Delta\epsilon_c$=0.94, corresponding
to $\Gamma_c^{LR}=2.15$.  
Figure~\ref{SHamp} shows $<\psi^2>$ as a function of $\epsilon$ for SH
simulations, and as a function of $\Gamma$ for MD simulations.
The fit shows good agreement in $\left<\psi^2\right>$ below
$\epsilon=0$ (Fig.~\ref{SHcomp}). 
Although the parameters are fit only in the range
$1.7\leq\Gamma\leq\Gamma_c^{LR}$, 
agreement is
reasonable even for
$\Gamma>\Gamma_c^{LR}$.

The three parameter fit not only allows for agreement in $\left<\psi^2\right>$,
but also matches the measure of order $P_{max}$ in the SH model to
that found
in MD simulation (Fig.~\ref{SHangle}).  
In both MD and SH
simulation, below the critical value of long range order, the fluctuations
are disordered, leading to a small value in $P_{max}$.  When $\Gamma$
crosses the critical
value, $P_{max}$ jumps up significantly, and the observed patterns are ordered
stripes.  Below the onset of stripes, when the fluctuations are constantly
shifting and
changing, there is significant uncertainty in finding the
value of $P_{max}$, as seen by the noisy curve on the plot.  Above this onset,
however, the standing waves produce stable stripes, and $P_{max}$ plateaus and
remains quite constant, with good
agreement between MD and SH simulations.
Finally, the mean field onset $\Gamma_{c}^{MF}=1.965\pm0.007$ predicted by
this fit agrees remarkably well with  
the critical value $\Gamma_{c}=1.955\pm0.005$ found in our simulations of
Navier-Stokes order continuum equations.

\subsection{Effect of changing wavelength on strength of noise}\label{section-wavelength}

If the noise effects arise from the finite
 particle number in MD, we may expect that this effect will
 decrease in
 systems in which there are more particles per wavelength of pattern.
 Since the
number of particles in a volume $\lambda^3$ increases with increasing
wavelength, we investigate the effect of changing frequency on the onset of
 patterns in
MD simulations.  For cells of horizontal extent $168\sigma\times10\sigma$,
layers shaken with a frequency $f^{*}=0.25$ form peaks with a dominant
wavelength $\lambda=42\sigma$, which is twice the wavelength found for
patterns investigated at $f^{*}=0.4174$ (see
Fig.~\ref{dispersion}).

\begin{figure}
\begin{center}
\scalebox{.5}{\includegraphics{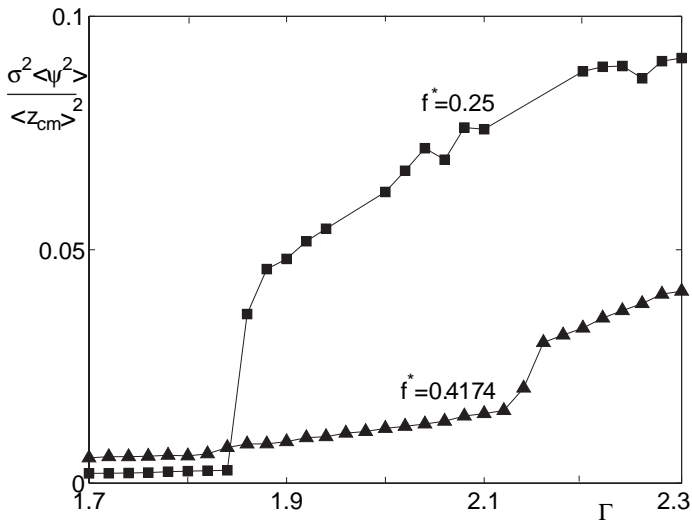}}
\end{center}
\caption{\label{noisefreqs} 
Comparison growth of $\left<\psi_{SH}^2\right>$ normalized by the mean center
of mass height of the layer $\sigma^2\left<\psi^2\right>/\left<z_{cm}\right>^2 =
\left<\left(z_{cm}-\left<z_{cm}\right>\right)^2\right>/\left<z_{cm}\right>^2$
for MD simulations with $f^*=0.25$ (squares) and $f^{*}=0.4174$ (triangles).
The lower frequency displays much smaller
fluctuations
below the onset of patterns than does the higher frequency.
}
\end{figure}

We examine layers shaken at $f^{*}=0.25$ in cells of size
$L_x=L_y=2\lambda=84\sigma$, while holding constant
layer depth $H=5.4$ and restitution coefficient $e=0.70$.  We vary
$\Gamma$ through the same range $1.7\leq\Gamma\leq2.3$ investigated for
$f^{*}=0.4174$ earlier in this paper.  
Figure~\ref{noisefreqs} shows the growth of $\left<\psi_{SH}^2\right>$ normalized
by the mean center
of mass height of the layer squared $\sigma^2\left<\psi^2\right>/\left<z_{cm}\right>^2 =
\left<\left(z_{cm}-\left<z_{cm}\right>\right)^2\right>/\left<z_{cm}\right>^2$
for MD simulations with frequencies $f^{*}=0.25$ and $f^{*}=0.4174$.

The lower frequency ($f^{*}=0.25$) exhibits a much sharper jump in $\left<\psi_{SH}^2\right>$
than that seen at $f^{*}=0.4174$.  Below this onset, the curve is
much flatter for $f^{*}=0.25$, while at $f^{*}=0.4174$,
the curve increases much more gradually through onset.  Proportionally
smaller fluctuations compared to pattern amplitude 
is consistent with lower noise strength for $f^{*}=0.25$
than that found for $f^{*}=0.4174$.  In addition, the
rapid growth of peaks and valleys occurs at a smaller value of
$\Gamma$ for $f^{*}=0.25$, corresponding to an onset even below the mean
field onset $\Gamma_c^{MF}$ for the larger frequency.

\begin{figure}

\hspace{0cm}
\subfigure{\label{GAM1.7lowf}\scalebox{.27}{\includegraphics{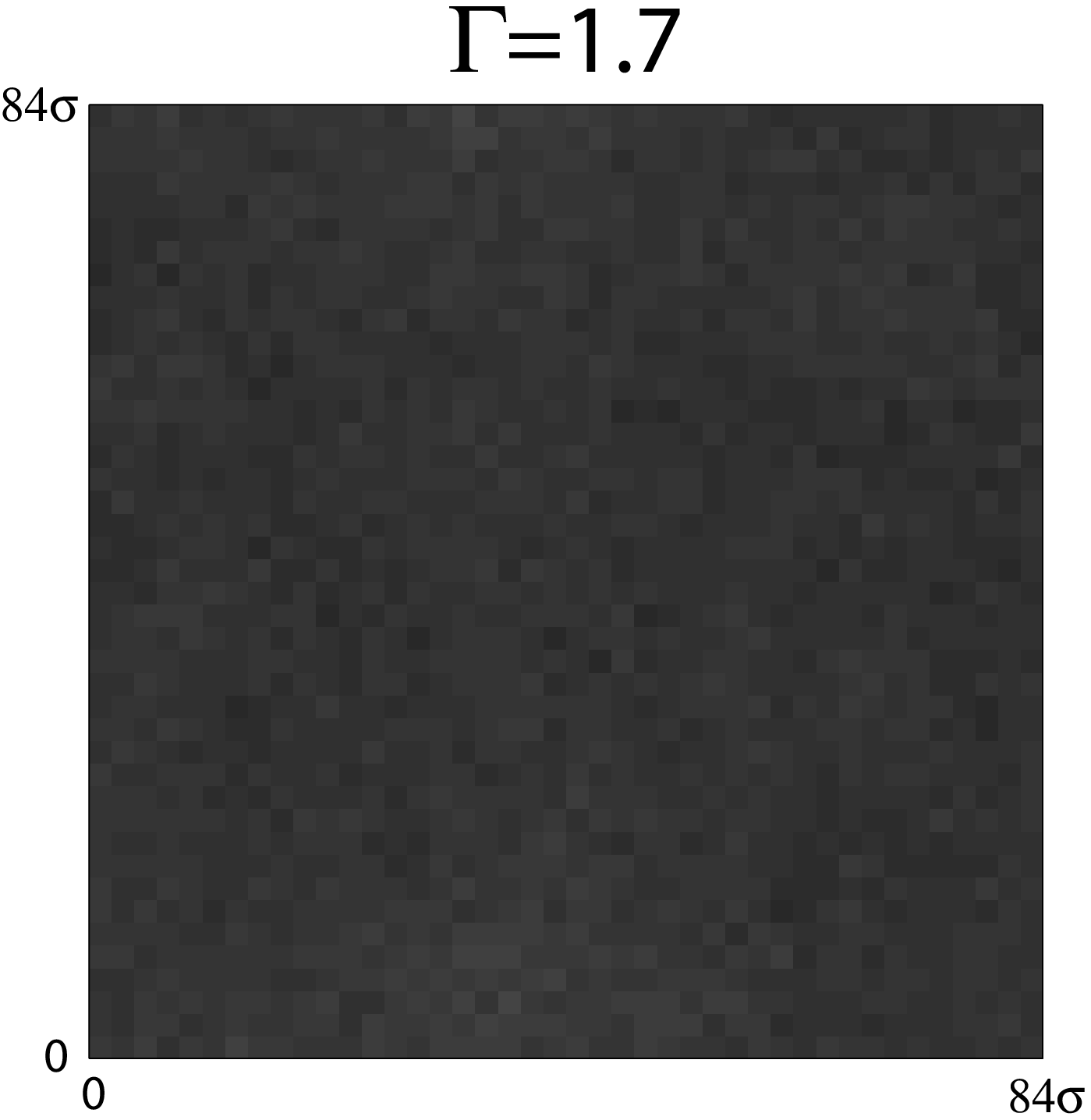}}}
\hspace{-0cm}
\subfigure{\label{GAM2.2lowf}\scalebox{.27}{\includegraphics{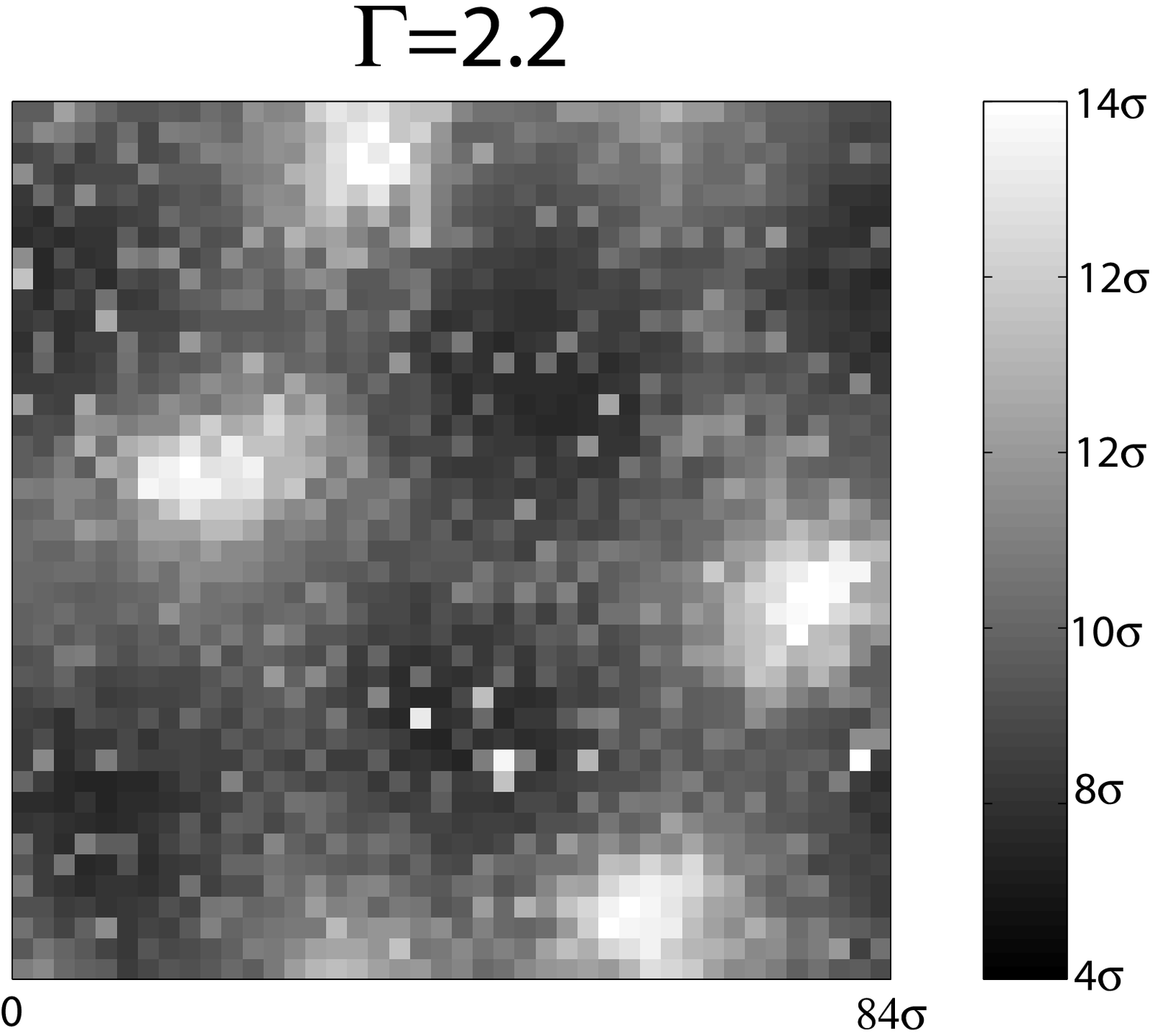}}}\\
\caption{\label{GAMlowf} 
An overhead view of the layer of grains from MD simulations at
$f^{*}=0.25$, for $\Gamma=1.7$ and $\Gamma=2.2$. Note how much less
noise there is below onset here ($\Gamma=1.7$) compared to the results for
$f^{*}=0.4174$ in
Fig.~\ref{GAMcomparison}.  The images show the center of mass height
$z_{cm}(x,y)$ of
the layer as a function of location in the box.  These MD simulations use a
cell which is
$L_{x}=L_{y}=84\sigma$ in the
horizontal directions.  Peaks corresponding to
large $z_{cm}$ are shown in white, while valleys corresponding to
small $z_{cm}$ are shown in black.  
The grayscale for both images is given on the right.
}
\end{figure}

We follow the same procedure as for $f^{*}=0.4174$ to fit the
data from MD simulation to the Swift-Hohenberg model.  
We note that for frictional
particles, square patterns are formed for $f^{*}=0.25$; in the
absence of friction, peaks and valleys remain disordered, and no regular
square lattice forms in experiments or MD simulations \cite{goldman03,
  moon03} (see Fig.~\ref{GAMlowf}).
Thus the onset of long range order is ill defined in this
case.  However, this lower frequency exhibits a much sharper onset in the
growth of $\left<\psi_{SH}^2\right>$, which is used to find $\Delta\epsilon_c$.
The best fit yields a noise term $F=\left(4\pm 3\right) \times 10^{-4}$,
and a mean field onset of $\Gamma_{c}^{MF}=1.85\pm0.01$.  Our hydrodynamic
simulations find the flat layer becomes unstable at
$\Gamma_{c}=1.84\pm0.01$, which again agrees well with the mean field onset
found from the fit to the SH model.

The noise strength at $f^{*}=0.4174$ is approximately 30 times
larger than the noise strength at $f^{*}=0.25$.  This leads to
qualitatively different behavior of $\left<\psi_{SH}^2\right>$ near onset,
yielding a smoother curve for the higher frequency and a sharper onset for
lower frequency.  Finally, the onset is barely delayed for the lower
frequency, with $\Delta\epsilon_c=0.01$ for $f^{*}=0.25$, as compared to 
 $\Delta\epsilon_c=0.10$ for $f^{*}=0.4174$.

Thus a change in frequency which increases the wavelength at onset by a
factor of $2$ decreases the amount of noise by a factor of $30$.  
For Rayleigh-B\'{e}nard convection in ordinary fluids, the
functional dependence of $F$ on 
$n$, ${\bf u}$, $T$, and $\lambda$ is known \cite{hohenbergswift92,
  vanbeijeren}.  However, this dependence is not known for oscillated granular
layers.  Future investigation of the dependence of $F$ on shaking parameters
$f^*$, $\Gamma$, and $H$, or on  hydrodynamic variables $n$, ${\bf u}$, $T$
in experiment and MD
simulations may provide information on the dependence of the noise strength
$F$ that can be used in continuum simulations.

\section{Conclusions}

We have shown that continuum simulations can describe important
aspects of pattern formation in granular materials.  
For a nondimensional frequency $f^*=0.4174$, both MD and continuum simulations
of granular materials form stripe patterns of the same wavelength above a
critical value $\Gamma>\Gamma_c$, and display no stripes for
$\Gamma<\Gamma_c$.  Further, the two simulations yield the same dependence
of wavelength on frequency.  These
wavelengths agree with the dispersion relation found experimentally
for frictional particles.

The effect of fluctuations has been examined in simulations of the
Swift-Hohenberg model.  The results deduced for the mean field onset in MD
simulations agree well with the actual onset in 
continuum simulations for both $f^{*}=0.4174$ and $f^*=0.25$.

We find the strength of the noise to be 
$F=\left( 1.2\pm0.2 \right) \times 10^{-2}$ for stripes at $f^{*}=0.4174$,
and  $F=\left( 4\pm3 \right) \times 10^{-4}$ for disordered squares at
$f^{*}=0.25$.  
The value determined in an experiment for
a slightly shallower granular layer at $f^{*}=0.28$ was $F=3.5 \times
10^{-3}$ \cite{goldman2004}, which is within the range of noise values
obtained in this investigation.
The smallest noise strength found for our granular system is
comparable to the largest noise strength found thus far
in experiments in ordinary fluids, which obtained $F=7.1 \times 10^{-4}$ in
measurements near the critical point, while
values typical for convection are closer to $10^{-9}$ \cite{oh}.
For $f^{*}=0.4174$, the noise is strong
enough to delay onset of long range patterns by $10\%$ in MD simulation,
and influences strongly the behavior of the system even more than
20\% below this onset.
Thus noise plays an important role in granular media near the onset of
patterns.  

This study indicates that hydrodynamic theory holds promise for
investigating and understanding pattern formation in granular flows.
However, quantitative comparisons between continuum theory and experiment
will require the addition of noise terms into the equations.  
The addition of noise would be an important step towards
using the powerful tools of hydrodynamic theory to investigate problems of
pattern formation in granular materials.  

The absence of
friction in these simulations restricts our investigation to stripe
patterns.  Simulations without friction have not yielded the square and
hexagonal patterns seen in experiments with frictional particles \cite{moon03}.
Further research into pattern
formation using continuum simulations should investigate the most
effective way to incorporate friction between particles into the continuum
simulations and should examine how the strength of friction in the simulation
affects pattern formation in the system.

\begin{acknowledgments}
We thank Daniel I. Goldman, W. D. McCormick, Sung Joon Moon, and Erin
C. Rericha for helpful
discussions.  This work was supported by the Engineering Research Program
of the
Office of Basic Energy Sciences of the Department of Energy (Grant
DE-FG03-93ER14312).
\end{acknowledgments}


\end{document}